\documentclass[aps,twocolumn,showpacs,prl,superscriptaddress,unsortedaddress]{revtex4}
\usepackage{epsf}
\usepackage{graphicx}
\usepackage{tabularx, booktabs, makecell, amsmath}
\usepackage{dcolumn}
\usepackage{bm}

\newcommand{\TaS}{$2H$-TaS$_2$\,}
\newcommand{\NbSe}{$2H$-NbSe$_2$\,}

\newcommand{\TiSe}{$1T$-TiSe$_2$\,}

\newcommand{\Tcdw}{$\it{T}_{\text{CDW}}\,$}

\newcommand{\wbar}{$\overline{\omega}$\,\,}

\begin{document}
     
     \title{ Spectroscopic fingerprints of many-body renormalizations in \TiSe } 
\author{J. Zhao}
\affiliation{Department of Physics, University of Virginia, Charlottesville, Virginia 22904}
\author{K. Lee}
\affiliation{Department of Physics, Ohio State University, Columbus, Ohio 43210}
\author{J. Li}
\affiliation{Department of Physics, University of Virginia, Charlottesville, Virginia 22904}
\author{D. B. Lioi}
\affiliation{Department of Physics, Drexel University, Philadelphia, Pennsylvania 19104}
\author{D. J. Gosztola}
\affiliation{Center for Nanoscale Materials, Argonne National Laboratory, Argonne, Illinois 60439, USA}
\author{G. P. Wiederrecht}
\affiliation{Center for Nanoscale Materials, Argonne National Laboratory, Argonne, Illinois 60439, USA}
\author{G. Karapetrov}
\affiliation{Department of Physics, Drexel University, Philadelphia, Pennsylvania 19104}
\author{N. Trivedi}
\affiliation{Department of Physics, Ohio State University, Columbus, Ohio 43210}
\author{U. Chatterjee}\thanks{uc5j@virginia.edu (U.C.)}
\affiliation{Department of Physics, University of Virginia, Charlottesville, Virginia 22904}

\date{\today}

\begin{abstract}
We have employed high resolution angle resolved photoemission spectroscopy (ARPES) measurements to investigate many-body renormalizations of the  single-particle excitations in \TiSe. The energy distribution curves of the ARPES data reveal intrinsic peak-dip-hump feature, while the electronic dispersion derived from the momentum distribution curves of the data  highlights, for the first time, multiple kink structures. These are canonical signatures of a coupling between the electronic degrees of freedom and some Bosonic mode in the system. We demonstrate this using a model calculation of the single-particle spectral function at the presence of an electron-Boson coupling. From the self-energy analysis of our ARPES data, we discern some of the critical energy scales of the involved Bosonic mode, which are $\sim$15 and 26 meV. Based on a comparison between these energies and the characteristic energy scales of our Raman scattering data, we identify these Bosonic modes as Raman active breathing (${\text{A}}_\text{1g}$) and shear (${\text{E}}_\text{g}$) modes, respectively. Direct observation of the band-renormalization due to electron-phonon coupling increases the possibility that electron-phonon interactions are central to the collective quantum states such as Charge density wave (CDW) and superconductivity in the compounds based on \TiSe.
\end{abstract}

\maketitle

\section{Introduction}
 The origin of various emergent phenomena in the solid state systems, such as superconductivity in cuprate high temperature superconductors (HTSCs) \cite{CUPRATE_REFERENCE1, CUPRATE_REFERENCE2, CUPRATE_REFERENCE3}, unusual mass renormalization in heavy fermion compounds \cite{HEAVY_FERMION_REFERENCE1, HEAVY_FERMION_REFERENCE2}, and colossal magnetoresistance (CMR) in manganites \cite{CMR_REFERENCE1, CMR_REFERENCE2}, is rooted to the many-body interactions. The electron-phonon (el-ph) coupling is a prominent member of the vast family of many-body interactions that are observed in correlated electron systems \cite{Mahan Book}. In the framework of the Bardeen-Cooper-Shrieffer (BCS) theory \cite{BCS_Theory}, the electron-electron pairing in conventional superconductors is mediated by the el-ph coupling. So is the case for the electron-hole pairing in majority of the charge density wave (CDW) systems \cite{Gruner}. Therefore, an in-depth understanding of the el-ph coupling is pivotal to interpret as well as to manipulate physical properties of the layered transition metal dichalcogenides (TMDs), where superconductivity and  charge density wave (CDW) are ubiquitous.  
 
 \TiSe, a widely studied TMD material, undergoes a second-order phase transition from a semimetal/semiconductor \cite{TISE2_SEMIMETAL_REF2, TISE2_SEMIMETAL_REF3, TISE2_SEMICONDUCTOR_REF1, TISE2_SEMICONDUCTOR_REF2} to a commensurate CDW state below the transition temperature (\Tcdw) $\sim$ 200 $K$ \cite{MONCTON_NS}. It has been shown that \Tcdw of \TiSe can be suppressed to zero either by chemical intercalation \cite{CAVA_NP, PD_CAVA}, or by strain engineering \cite{TISE2_PRESSURE}. In each case, the superconductivity emerges in a dome-shaped region of the corresponding phase diagram, reminiscent of those of the HTSCs \cite{CUPRATE_REFERENCE1, CUPRATE_REFERENCE2, CUPRATE_REFERENCE3} and heavy fermion compounds \cite{HEAVY_FERMION_REFERENCE1, HEAVY_FERMION_REFERENCE2}. Despite extensive investigations, the mechanisms of the CDW order in pristine \TiSe and the superconductivity in Cu-intercalated \TiSe are topics of ongoing controversies. 

Two types of models \cite{EXCITONIC_INSULATOR_1, EXCITONIC_INSULATOR_2, JT_HUGE, JT_WANGBO, BERATHING_ARXIV}, which provide diverging explanations, have been proposed as possible candidates for the CDW order in \TiSe. In the first type of models, the long-range CDW order is triggered by the condensation of excitons at \Tcdw \cite{EXCITONIC_INSULATOR_1, EXCITONIC_INSULATOR_2, PETER_SCIENCE}. The second type of models propose the CDW transition to be some variant of a Jahn-Teller-like instability, which occurs due to strong electron-phonon coupling in the system \cite{JT_HUGE, JT_WANGBO}. Previously, the results of a number of ARPES measurements have been interpreted using the excitonic condensation model \cite{AEBI_1, AEBI_2,AEBI_3,AEBI_4, AEBI_5, AEBI_6}. Recent scanning tunneling microscopy studies \cite{AEBI_STM1, AEBI_STM2} and ultrafast spectroscopic measurements \cite{KYLE_ULTRAFAST_OPTICS}, however, highlight the significance of Jahn-Teller-like distortions and hence, that of electron-phonon interactions to the CDW order in \TiSe. Similarly, as to the superconductivity in Cu-intercalated \TiSe samples, there are two contrasting views of the superconducting glue. The first one, which relies on the scenario of a phase competition between the superconducting and CDW orders, suggests superconductivity to be stabilized by quantum fluctuations of the CDW order above certain critical concentration of the Cu atoms for which the CDW order disappears \cite{TiSe2_HASAN, CAVA_NP}.
According to the second hypothesis, a combination of enhanced el-ph coupling and increased density of states at the chemical potential of the system gives rise to the superconductivity in the samples with high concentration of Cu atoms \cite{TiSe2_FENG, TISE2_DFT2}. Given all these, a direct investigation of the el-ph coupling in \TiSe-based compounds will be highly desirable. An important step towards this direction will be to first explore the direct signatures of the el-ph coupling in the parent compound of \TiSe. 

There are comprehensive theoretical works on various aspects of phonons in \TiSe. For instance, Motizuki and coworkers \cite{Theory_TiSe2_1, Theory_TiSe2_2, Theory_TiSe2_3}, developed a general picture of the lattice distortions in TMDs including \TiSe. Recent first-principles calculations \cite{TISE2_DFT2} reported that the CDW transition in pristine  as well as the  superconductivity in pressurized  \TiSe samples can entirely be determined by the electron- phonon interaction. Additionally, it has been suggested that the electron-phonon interactions must be taken into account \cite{TISE2_BISHOP} to fully understand the origin of  the chiral nature of the CDW state in \TiSe.
On the experimental front, phonon density of states and phonon softening  have been thoroughly probed by X-ray \cite{STEPHAN_PRL, HOLT_PRL, FRANK_PRL, FRANK_PRB} and Raman scattering experiments \cite{RAMAN_REF1, RAMAN_REF2}. However, direct examination of the el-ph coupling using these techniques is complicated. 

A straightforward way to investigate the subtle details of the coupling between a Bosonic mode  and the electronic excitations of a solid is to concentrate on its single-particle self-energy \cite{Mahan Book}. The net effect of such a coupling in a system is anticipated to be a renormalization of the various attributes of its quasiparticle. In principle, ARPES measurements from a solid can be used to gain knowledge of its self-energy. A manifestation of the renormalizations due to an electron-mode coupling is the appearance of a discontinuity, i.e., the so-called $kink$, in the renormalized dispersion. This can be understood as follows: the dispersion close to the chemical potential becomes flatter due to an enhancement in the effective mass of the quasiparticles, while the dispersion sufficiently away from the chemical potential maintains its bare form. The energy scale of the mode and its coupling strength can approximately be evaluated from the location and the strength of the kink, respectively. Indeed, a large body of works  have been devoted to the study of the dispersion kinks in different TMDs \cite{TMD_KINK1, TMD_KINK2, TMD_KINK3, TMD_KINK4, UC_PRB, UC_JMC}.  Strikingly, such a study on \TiSe is lacking. This motivates the present self-energy analysis of the ARPES data from \TiSe, where we make the first observation of multiple kink structures in the electronic dispersion because of el-ph coupling.

 \section{Experimental Details}
 
We have conducted ARPES measurements on \TiSe single crystals  using 21.2 eV Helium-I line of a discharge lamp combined with a Scienta R3000 analyzer at the University of Virginia, as well as 24 and 43 eV synchrotron light equipped with a Scienta R4000 electron analyzer at the SIS beamline of the Swiss Light Source, Paul Scherrer Institute, Switzerland. The energy and momentum resolutions were approximately 8-20 meV and 0.0055 \AA$^{-1}$ respectively. Single crystals were cleaved in situ to expose a fresh surface of the crystal for ARPES measurements. Samples were cooled using a closed cycle He refrigerator and the sample temperatures were monitored using a silicon diode sensor mounted close to the sample holder. During each measurement, the chemical potential ($\mu$) of the system was determined by analyzing ARPES data from a polycrystalline gold sample in electrical contact with the sample of interest. High quality single crystals of \TiSe were grown using the standard iodine vapor transport method and  the samples were characterized using X-ray diffraction, energy dispersive X-ray spectroscopy (EDS), and electrical resistivity measurements. Temperature-dependent Raman scattering measurements were performed at the Center for Nanoscale Materials at Argonne National Laboratory, using the Renishaw In Via Raman microscope with a 514 nm argon ion laser source and a $\sim1.5\mu$m diameter spot size. The spectrometer is equipped with variable temperature cell capable of operating between 80 and 500K. All the experiments were conducted in the presence of ultra-high pure nitrogen exchange gas at normal pressure.

 \begin{figure}
\centering
\includegraphics[width=0.48\textwidth]{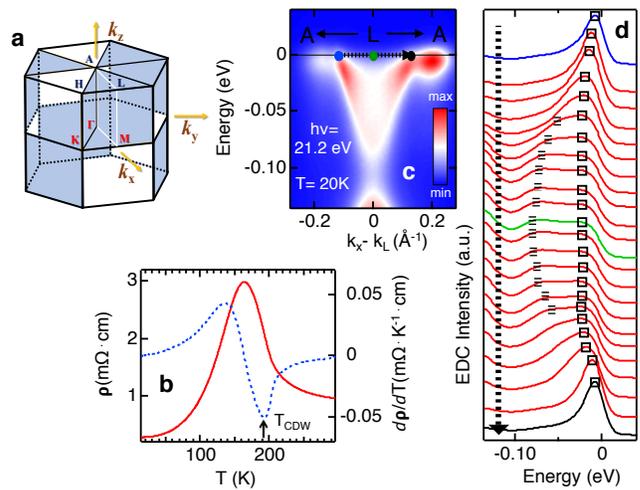}
\caption{(a) A schematic layout of the normal state Brillouin zone of \TiSe containing high-symmetry points. (b) Temperature dependence of the in-plane electrical resistivity ($\rho$) of \TiSe. \Tcdw\ is is determined from the minimum (pointed by the black arrow) of  $\displaystyle\frac{d\rho}{dT}$ vs $T$ plot (black dashed line). (c) EMIM of the conduction band around L point. (d) EDCs between the two Fermi crossings marked by the black and blue dots in (c). These EDCs are offset for visual clarity. Peak-dip-hump structure of these EDCs are clearly visible in. Peaks are shown by the open squares, while the humps are by dashed lines. The ARPES data has been recorded with $h\nu=21.2$ eV at 20K.}
\end{figure}

\section{Results}
\subsection{A. Intrinsic Peak-dip-hump structure of the energy distribution curves}

We start with a schematic layout of the normal state three-dimensional Brillouin zone of \TiSe in Fig. 1(a), which shows various high-symmetry points. \Tcdw $\sim200$K can be verified from electrical resistivity vs temperature plot in Fig. 1(b). In order to explore the signatures of many-body interactions in the system, we will focus on the line shape analysis of the energy distribution curves (EDCs) around L point. An EDC is ARPES intensity as a function of electronic energy at a specific momentum location. In Fig. 1(d), we present a stack of EDCs at 20K, which are associated with an ARPES energy-momentum intensity map (EMIM) around L point as shown in Fig. 1(c). An EMIM is ARPES intensity as a function of one of the in-plane momentum components and electronic energy ($\overline{\omega}$) referenced to $\mu$, while the remaining in-plane momentum component is fixed. These EDCs clearly display two-peak features, commonly known as the peak-dip-hump (PDH) structure. 

If the PDH structure of the EDCs is found to be intrinsic, it can be ascribed to a nontrivial many-body interaction, a coupling of the electrons to some Bosonic mode, for instance. To examine this, we analyze photon energy ($h\nu$) dependence of the ARPES data in Fig. 2. In this context, the EMIM in Fig. 1(c) is recorded with $h\nu=21.2$ eV. Two other EMIMs are shown Figs. 2(a) ($h\nu=$ 24 eV) and 2(b) ($h\nu=$ 43 eV). EDCs constructed from Figs. 2(a) and 2(b) are exhibited in Figs. 2(c) and 2(d), respectively. The PDH structure of the EDCs is clear in each case. We have also checked that the variations of intensities of the peaks and humps of the EDCs at equivalent momenta scale together reasonably well with changing $h\nu$. 
Collectively, these observations lead us to conclude that the PDH line shape of \TiSe represents a single electronic state governed by a coupling between electronic degrees of freedom and some Bosonic mode. This is further illustrated by incorporating a model calculation of the spectral function in section $\bf{B}$. Moreover, a detailed discussion on the relevant many-body interactions and the nature of the involved Bosonic mode is presented in section $\bf{C}$ by adopting the self-energy analysis of our ARPES data.
\begin{figure}
\centering
\includegraphics[width=0.48\textwidth]{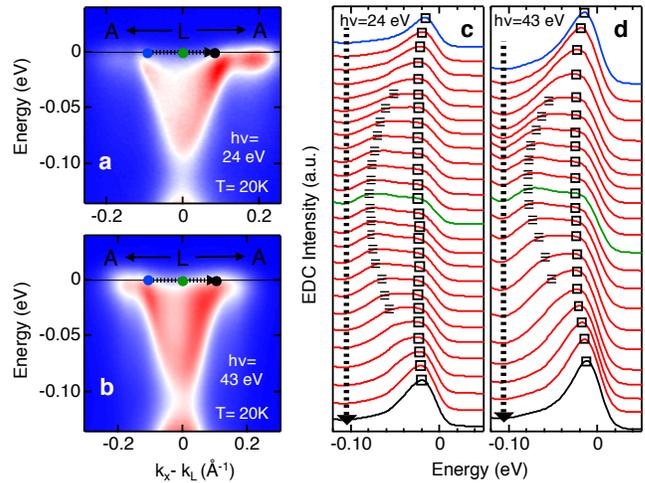}
\caption{ EMIMs, similar to that in Fig. 1(c), are displayed in (a) and (b) with $h\nu=24$ and 43 eV, respectively. Stacks of EDCs corresponding to (a) and (b) are shown in (c) and (d), respectively. It can be observed that the PDH structure of the EDCs is independent of photon energy.}
\end{figure}
\subsection{B. Model calculation of PDH structure based on a coupling between electrons and an Einstein mode}
\begin{figure}
\centering
\includegraphics[width=0.48\textwidth]{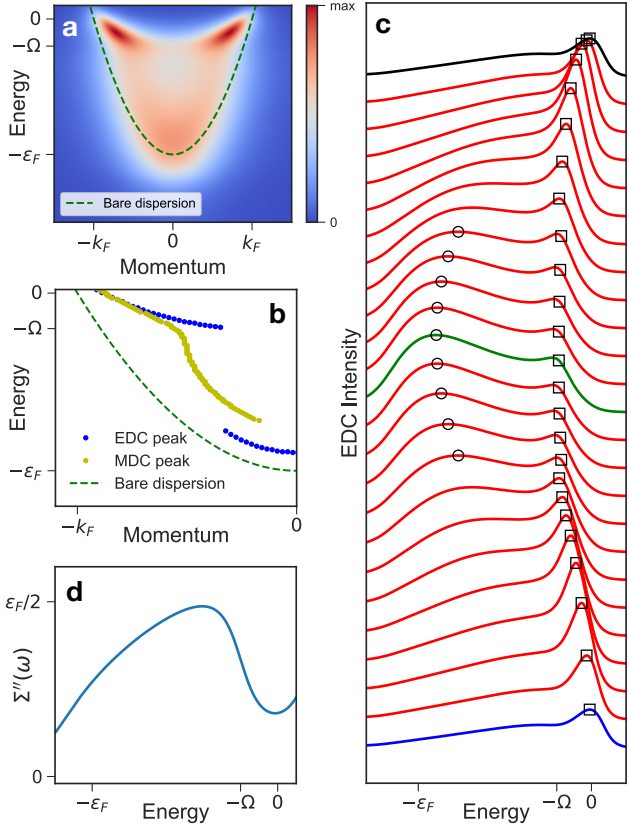}

\caption{(a) Intensity map of the calculated spectral function $A(\mathbf{k}, \omega)$ of electron coupled to a Bosonic mode, multiplied by the Fermi-Dirac function ($T/\epsilon_F=0.05$).
(b) Dispersions extracted from the calculated $A(\mathbf{k}, \omega)$.
Blue and yellow markers respectively represent peaks positions of EDC and MDC, whose dispersions show a kink.
Red dashed line marks the bare dispersion.
c) EDCs between $-k_F$ and $k_F$ of the bare dispersion.
Blue, green, and black lines represent EDC at $k=-k_F, 0$, and $k_F$, respectively.
Black squares mark the locations of the ``peaks'' as defined by the maximum curvature in EDC between $0$ and $-2\Omega$, and black circles mark the locations of the ``humps'' as defined by the local maximum in EDC below $-2\Omega$.
(d) Imaginary part of self-energy.}
\end{figure}

For comparison, we have calculated the spectral function $A(\mathbf{k}, \omega)$ of electrons coupled to a Bosonic mode.
The single-particle electron self-energy $\Sigma(\mathbf{k}, \omega)$ from mode-coupling is
\[
  \Sigma(\mathbf{k}, \omega) =- g^2 \int \!\mathrm{d}\omega' \mathrm{d}^{d}k' \;
    G(\mathbf{k}', \omega')
    D(\mathbf{k} - \mathbf{k}', \omega - \omega'),
\]
where $G$ and $D$ are the propagators of the electron and the Bosonic mode, and $g$ the coupling strength.
In Fig. 3, we show the calculated spectral function and self-energy.
The electron is assumed to have a three-dimensional quadratic dispersion $\epsilon_\mathbf{k} / \epsilon_F = (k/k_F)^2-1$, and the Bosonic mode to have an Einstein dispersion with energy $\Omega/\epsilon_F = 0.2$.
We used the coupling strength $(g/\epsilon_F)^2 = 0.4$, and the intrinsic scattering rate $\eta/\epsilon_F = 0.1$.
A kink structure in the electronic dispersion is clearly visible in Figs.  3(a) and 3(b).

With Einstein dispersion of the Bosonic mode, and momentum-independent coupling constant, the resulting electron self-energy also is momentum-independent, and only depends on the energy: $\Sigma(\mathbf{k}, \omega) = \Sigma(\omega)$.
At energies smaller than the mode energy $\Omega$, the imaginary part of the self-energy  $\Sigma''(\omega)$ should vanish, as the Bosonic propagator becomes purely virtual.
In Fig. 3(d), we show the form of $\Sigma''(\omega)$ we have used in the calculation of $A(\mathbf{k}, \omega)$  in Fig. 3(a).
Although $\Sigma''$ remains nonzero between $\pm \Omega$ because of the intrinsic broadening $\eta$, it will vanish as $\eta \rightarrow 0^{+}$.
The EDCs of the calculated $A(\mathbf{k}, \omega)$, are shown in Fig. 3(c). Similar to those in Fig.1 and Fig. 2, these EDCs also display PDH structure.

\subsection{C. Kink structure in the electronic dispersion}
$\Sigma(\textbf{k}, \overline{\omega})$ is a complex-valued function of momentum and energy. Its real part ${\Sigma}^{'}(\textbf{k},\overline{\omega})$ contains information on the renormalizations of the bare electronic dispersion, while the imaginary part ${\Sigma}^{''}(\textbf{k},\overline{\omega})$ represents the single-particle lifetime at the presence of interactions \cite{Mahan Book}. Using ARPES data, $\Sigma^{'}$  and $\Sigma^{''}$ can, in principal, be directly obtained. This can be realized by noting that ARPES intensity $\text{I}(\textbf{k}, \overline{\omega})$ can approximately be written as follows:  $\text{I}(\textbf{k}, \overline{\omega}) \sim M(\textbf{k})\text{A}(\textbf{k},\overline{\omega})f(\overline{\omega})$, where (i) $f(\overline{\omega})$ is the Fermi-Dirac distribution function, (ii) $M(\textbf{k})$ is the dipole matrix element, (iii) the spectral function $\displaystyle \text{A}(\textbf{k}, \overline{\omega})=\frac{\Sigma^{''} (\textbf{k}, \overline{\omega})}{(\overline{\omega}-\epsilon_\textbf{k}-\Sigma^{'} (\textbf{k},\overline{\omega}))^2+{\Sigma^{''}(\textbf{k}, \overline{\omega})}^2}$ and (iv) $\epsilon_\textbf{k}$ is the bare electronic dispersion \cite{HUFNER_BOOK, JC_REVIEW, ZX_REVIEW, JHONSON_REVIEW}. Self-energy analysis from the data, however, becomes operationally manageable only when $\Sigma^{'}$  and $\Sigma^{''}$ are independent of $\textbf{k}$ or in certain cases with weak $\textbf{k}$ dependence.

In case of $\textbf k$-independent $\Sigma^{'}$, $\Sigma^{''}$, MDCs at various $\overline{\omega}$'s take simple Lorentzian line shape, at least in the vicinity of the Fermi momentum $k_F$ where $\epsilon_\textbf{k}$  can be approximated as follows:  $\epsilon_\textbf{k}\sim v^*_F(|\textbf{k}|-k_F)$ with $v^*_F$ being the renormalized Fermi velocity. 
The renormalized dispersion of an energy band can be determined by plotting the fitted peak positions of the corresponding MDCs as a function of \wbar.
The deviation of this renormalized dispersion from the bare dispersion provides a measure for $\Sigma^{'}(\overline{\omega})$ \cite{HUFNER_BOOK, JC_REVIEW, ZX_REVIEW, JHONSON_REVIEW}. Additionally, $\Sigma^{''}(\overline{\omega})$ can be quantified from the fitted peak widths $W(\overline{\omega})$ of the MDCs. The relation between $\Sigma^{''}(\overline{\omega})$ and $W(\overline{\omega})$ is as follows: $\displaystyle W(\overline{\omega})=\frac{\Sigma^{''}(\overline{\omega})}{v_F^*}$ \cite{HUFNER_BOOK, JC_REVIEW, ZX_REVIEW, JHONSON_REVIEW}. 
\begin{figure}
\centering
\includegraphics[width=0.48\textwidth]{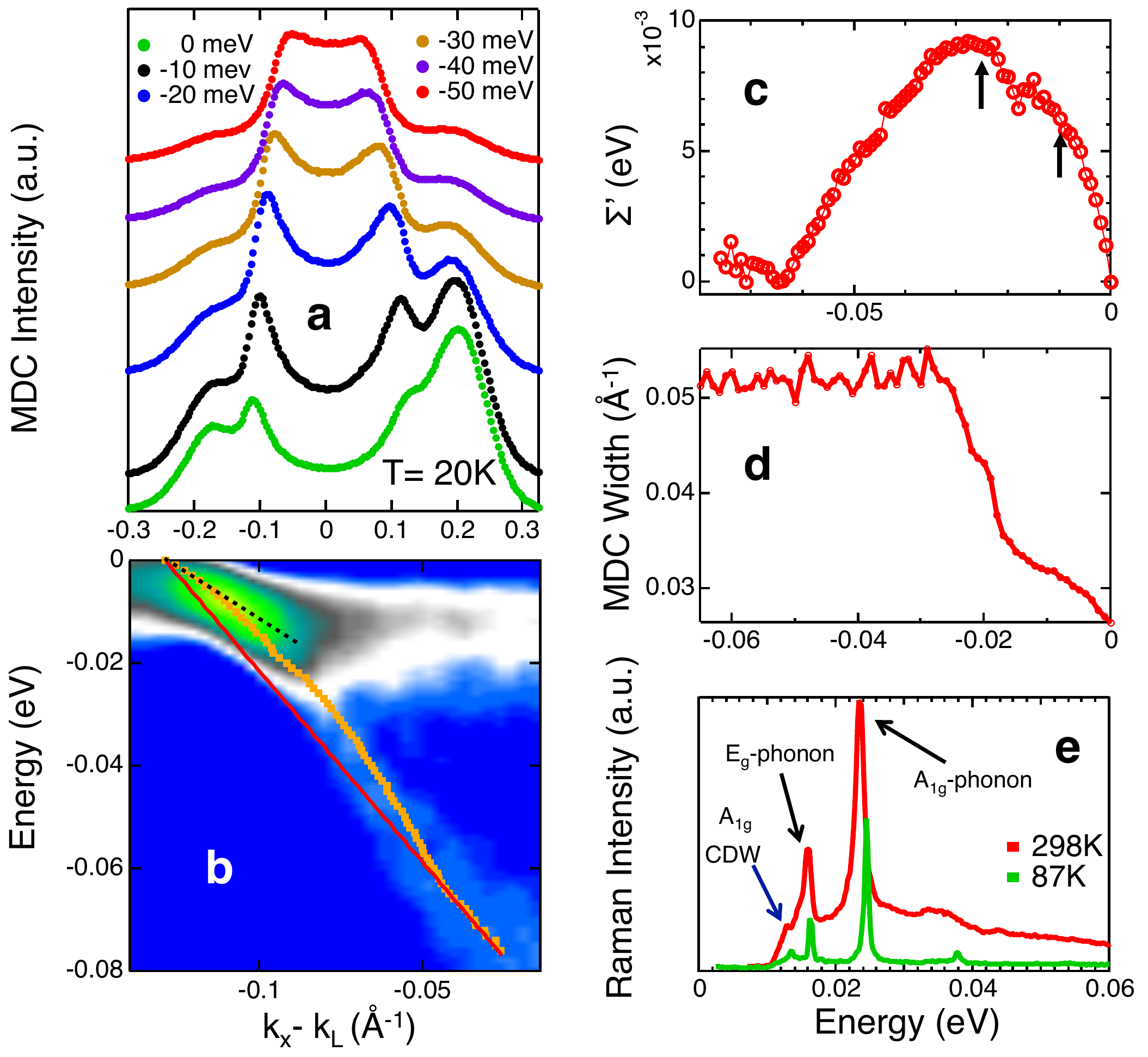}
\caption{(a) MDCs for a series \wbar's along the momentum line, shown by the black dashed arrow in Fig. 1(c). These MDCs are offset for visual clarity. Band dispersion (orange markers) from MDC analysis is superimposed on the second derivative of the energy-momentum intensity map of Fig. 1(c) with respect \wbar in (b). Approximated bare band dispersion is shown by the red line, whose slope determines $v_F^0$. The slope of the black dashed line determines $v_F^*$. Kinks in the renormalized dispersion can easily be detected. Here we have focussed on the left branch of the intensity map. We have checked that the right branch also gives similar result. $\Sigma^{'}(\overline{\omega})$ and $W(\overline{\omega})$ are displayed in (c) and (d), respectively. Black arrows in (c) point peaky structures of  $\Sigma^{'}(\overline{\omega})$. Comparing (c) and (d), it can be seen that the \wbar locations of the black arrows approximately match with the \wbar's, at which the slope of $W(\overline{\omega})$ changes. Note that $\Sigma^{'}(\overline{\omega})$ is directly proportional to $W(\overline{\omega})$. (e) Temperature evolution of Raman spectra of \TiSe single crystal. Raman data displays energy scales of the CDW amplitude mode, and breathing (${\text{A}}_\text{1g}$) and shear(${\text{E}}_\text{g}$) phonon modes.}
\end{figure}
 Fig. 4(a) presents MDCs for several values of \wbar along the momentum line marked by the black dashed arrow in Fig. 1(b)). In Fig. 4(b), we superimpose the dispersion curve on the second derivative of the EMIM with \wbar. A closer look at the dispersion curve in Fig. 4(b) further reveals that the band dispersion consists of multiple changes in slope, commonly referred as the kinks. Similar kink features in the electronic dispersion have also been observed in a wide array of solid state systems, including various $2H$-polytypes of TMDs \cite{TMD_KINK1, TMD_KINK2, TMD_KINK3, TMD_KINK4, UC_PRB, UC_JMC}, metallic systems \cite{METAL_KINK1, METAL_KINK2}, conventional superconductors \cite{ADAM_MgB2}, manganites \cite{DAN_LSMO_PRL}, cuprate high temperature superconductors \cite{JC_REVIEW, ZX_REVIEW, JHONSON_REVIEW}, and pnictide high temperature superconductors \cite{HASAN_KINK_PNICTIDE}.

\subsection{D. Identity of the Bosonic mode }
Typically, the  presence of a kink in the electronic dispersion is interpreted as a fingerprint of electronic scatterings from a Bosonic mode of the system \cite{JC_REVIEW, ZX_REVIEW,Mahan Book}. In order to address the identity of the mode in the present case, we take resort to the self-energy analysis of our ARPES data. 
The knowledge of the bare band dispersion is necessary for evaluating $\Sigma^{'}(\overline{\omega})$ from the data. This is approximated by a straight line, which follows the high binding energy part of the MDC-derived dispersion and it passes through $k_F$. Similar approximation has been used for other systems, too \cite{TMD_KINK3, JC_REVIEW, ZX_REVIEW, JHONSON_REVIEW, HASAN_KINK_PNICTIDE, DAN_LSMO_PRL, ADAM_MgB2}. 
 We quantify $\Sigma^{'}(\overline{\omega})$ by subtracting the approximated bare band dispersion from the measured one. Additionally, $\Sigma^{''}(\overline{\omega})$ can be obtained from $W(\overline{\omega})$. $\Sigma^{'}(\overline{\omega})$ and $W(\overline{\omega})$ are plotted in Figs. 4(c) and 4(d), respectively. A closer look at Fig. 4(c) reveals that $\Sigma^{'}(\overline{\omega})$ is associated with a number of peaks. The prominent peaks-energies in the present case are $\sim$15 and 26 meV, which agree well with the Raman active breathing (${\text{A}}_\text{1g}$) and shear(${\text{E}}_\text{g}$) modes, respectively (Fig. 4(e)).
 These are consistent with other Raman Scattering measurements on the system \cite{RAMAN_SCATTERING_TISE2_1, RAMAN_SCATTERING_TISE2_2, RAMAN_REF1, RAMAN_SCATTERING_TISE2_4, RAMAN_SCATTERING_TISE2_5}. Therefore, it would be natural to conclude that the electron-phonon coupling is responsible for the renormalization of the electronic dispersion. It is worth mentioning that similar multiple kinks of phononic origin have been reported in ARPES studies of other TMDs, such as  \NbSe \cite{TMD_KINK3, TMD_KINK2} and \TaS \cite{UC_JMC}.
 
To correlate our MDC analysis with the electrical transport measurements of the system, we estimate electrical resistivity ($\rho$) using Drude formula:  $\displaystyle \rho=\frac{m^*}{ne^2\tau}$, where $m^*$ is the effective mass of the charge carriers, $n$ is the carrier-density and $\tau$ is the scattering time. We find $n\sim10^{20}\text{cm}^{-3}$ from our Hall measurements. Other two Drude parameters, namely $\tau$ and $m^*$, can be obtained from the MDC analysis \cite{YOSHIDA_TRANSPORT_ARPES}. An estimate for $\tau$ is as follows: $\displaystyle \tau\sim\frac{\hbar}{\Sigma^{''}(\overline{\omega}=0)}\sim23$ fs. Furthermore, $m^{*}$ can be written as: $\displaystyle m^*\sim(1+\lambda)m_e$, where $\lambda$ is the mass enhancement due to many-body interactions and $m_e$ is the electronic mass. Note that $\lambda$ can also be taken as a measure for the  electron-Boson coupling. To be precise, we should have used band-mass $m_{LDA}$ instead of $m_e$ in the previous expression for $m^{*}$. Given that $m_{LDA}$ is not expected to be significantly different from $m^*$ and we are only trying to have an approximate value for $\rho$, we use $m_e$  in the previous expression. Furthermore, $\lambda$ can be quantified from the following relation: $\displaystyle \lambda=\frac{v_F^0}{v_F^*}-1$, where $v_F^0$ is the bare Fermi velocity, i.e., the slope of the approximated bare dispersion, and  $v_F^*$ is the renormalized Fermi velocity, i.e., the slope of the renormalized dispersion at the chemical potential. From Fig. 4(b), we find that $v_F^*=0.54$ eV$\cdot${\AA} and $v_F^0=0.78$ eV$\cdot${\AA}. 
Finally, we obtain: $\rho\sim2.24$ m$\Omega\cdot$cm, which is in reasonably good agreement  with the experimentally found value of $\rho\sim0.4$ m$\Omega\cdot $cm (Fig. 1(b)) 
 

\section{Conclusions}
In summary, we report here the first observation of multiple kink structures due to electron-phonon coupling in the ARPES spectra of \TiSe around L point. Employing self-energy analysis of our data, we decipher the energy scales of the phonon modes---$\sim$ 15 meV and $\sim$ 26 meV. These energies match nicely with those of Raman active breathing (${\text{A}}_\text{1g}$) and shear (${\text{E}}_\text{g}$) phonon modes.
 Furthermore, the estimated value on the electron-phonon coupling of \TiSe $\sim 0.5$, which makes this system a moderately coupled one. Direct observation of the clear signature of  electron-phonon coupling from ARPES provides support to the theoretical models, in which the CDW transition in \TiSe is proposed to be triggered by electron-phonon interaction induced Jahn-Teller-like instability. 

~\\ \indent
{\bf Acknowledgments} \\
U.C. and N. T. acknowledge support from the National Science Foundation (NSF) under Grant No. DMR-1629237. G.K. acknowledges the support by the National Science Foundation under Grant No. ECCS-1408151.The use of the Center for Nanoscale Materials, an Office of Science user facility, was supported by the U.S. Department of Energy, Office of Science, Office of Basic Energy Sciences, under Contract No. DE-AC02-06CH11357.

\end{document}